\documentclass[runningheads]{llncs}
\usepackage{graphicx}
\usepackage{url}
\usepackage{tabularx}
\usepackage{color}
\usepackage[flushleft]{threeparttable}

\begin{document}

\title{SMaRt Blockchain Distributed Workflow Management}

\titlerunning{SMaRt Blockchain Distributed Workflow Management}

\author{Joerg Evermann\inst{1} 
\and
Henry Kim\inst{2}}

\authorrunning{J. Evermann and H. Kim}

\institute{Memorial University of Newfoundland, St. John's, Canada\\
\email{jevermann@mun.ca} \and
York University, Toronto, Canada\\
\email{hkim@york.ca}}

\maketitle

\begin{abstract}
Blockchain technology has been proposed as a new infrastructure technology for a wide variety of novel applications. Blockchains provide an immutable record of transactions, making them useful when business actors do not trust each other. Their distributed nature makes them suitable for inter-organizational applications. However, proof-of-work based blockchains are computationally inefficient and do not provide final consensus, although they scale well to large networks. In contrast, blockchains built around Byzantine Fault Tolerance (BFT) algorithms are more efficient and provide immediate and final consensus, but do not scale well to large networks. We argue that this makes them well-suited for workflow management applications that typically include no more than a few dozen participants but require final consensus. In this paper, we discuss architectural options and present a prototype implementation of a BFT-blockchain-based workflow management system (WfMS).


\end{abstract}

\section{Introduction}

Inter-enterprise business processes may include stakeholders in adversarial relationships, that nonetheless have to jointly complete process instances. Trust in the current state of a process instance and correct execution of activities by other stakeholders may be lacking. Blockchain technology can help in such situations by providing a trusted, distributed workflow execution infrastructure.

A blockchain cryptographically signs a series of blocks, containing transactions, so that it is difficult or impossible to alter earlier blocks in the chain. In a distributed blockchain, actors independently validate transactions, add them to the blockchain, and replicate the chain across different nodes. The independent and distributed nature of actors requires finding a consensus regarding the validity and order of transactions and blocks. In workflow execution, it is important that actors agree on the ''state of work'' as this determines the set of next valid activities in the process. Hence, it is natural to use blockchain transactions to describe workflow activities or workflow states. 

In contrast to prior work, which has focused on transaction ordering on proof-of-work blockchains, we examine the use of consensus protocols based on algorithms for Byzantine Fault Tolerance (BFT). Furthermore, we explore the architecture of a blockchain-based WfMS without smart contracts. We motivate both of these choices later in the paper. Event without the use of smart contracts, the blockchain remains essential as it provides independent validation of workflow activities, distribution, replication, and tamper-proofing to workflow execution.

Blockchain technology admits many different system designs, and WfMS can be implemented in many different ways on blockchain infrastructure. In this paper, we focus on the interface between the blockchain and the workflow engine and the architectural options available for the design of the system. 

\paragraph*{Contributions} We describe a prototype WfMS system as a proof-of-concept implementation for an architecture that has not yet seen attention in the literature. First, in contrast to earlier work (Sec.~\ref{sec:related}) we do not use smart contracts to implement model-specific workflow engines. We show that generic or existing workflow engines can be readily adapted to fit onto a blockchain infrastructure and that smart contracts are not required. Second, as recommended, but not implemented by \cite{ViriyasitavatHoonsopon2018blockchain}, we show how a BFT-based blockchain can be used as workflow management infrastructure. We describe the implementation of a blockchain-based WfMS that has served as our tool to investigate design choices, problems and solutions in this research area. While our prototype is an important demonstration of feasibility, our main contribution is in the identification and discussion of the different architectural choices, and highlighting the existence and feasibility of alternatives to smart contracts on proof-of-work blockchains in this research area.

The remainder of the paper is structured as follows. Section~\ref{sec:related} reviews related work on blockchain-based WfMS. We then describe the principles of distributed blockchains with a focus on BFT-based consensus (Sec.~\ref{sec:blockchains}). Section~\ref{sec:architecture} describes the architecture of our system and discusses design choices. Section~\ref{sec:prototype} presents our prototype implementation. The final Sec.~\ref{sec:discussion} discusses implications of BFT-based blockchain technology for WfMS and an outlook to future work.

\section{Related Work}
\label{sec:related}

This section discusses existing work in two research areas. The first subsection focuses on blockchain technology applied to workflow management; the second subsection focuses on blockchains that apply a BFT ordering mechanism.

\subsection{Workflow Management on Blockchains}

Blockchain-based workflow management has only recently received research attention \cite{mendling2018blockchains}. The main research challenges are around integration of blockchain infrastructure into WfMS and ensuring correctness and security of the workflow execution \cite{mendling2018blockchains}. A number of prototype implementations have been presented, focusing on the use of ''smart contracts''. A smart contract is a software application that is recorded and executed on the blockchain. This application ''listens'' for relevant transactions sent to it and executes application logic upon receipt of a transaction. For example, the widely used Ethereum\footnote{\url{https://ethereum.github.io/yellowpaper/paper.pdf}} blockchain has a Turing-complete virtual machine (VM) for smart contracts and compilers for different programming languages.

In a project driven by a financial institution, a prototype workflow implementation using smart contracts on the Ethereum blockchain offers digital document flow in the import/export domain \cite{fridgen2017implementation,fridgen2018cross}. The project demonstrates significantly lowered process cost, as well as increased transparency and trust among trading partners.

A blockchain-based workflow project in the real-estate domain \cite{hukkinen2017distributed}, also using the Ethereum blockchain and smart contracts, notes that the de-centralized nature of blockchains and the lack of a central agency will make it difficult for regulators to enforce obligations and responsibilities of trading partners.

A complete WfMS, including collaborative workflow modelling and model instantiation, uses models as contracts between collaborators \cite{harerdecentralized}. The system allows distributed, versioned modelling of private and public workflows, consensus building on versions to be instantiated, and tracking of instance states on the blockchain. The blockchain provides integrity assurance for models and instance states. The authors note that the usefulness of the approach is limited by block size limits on the blockchain and the latency of new blocks \cite{harerdecentralized}. 

Another implementation of blockchain-based workflow execution \cite{weber2016using,weber2016untrusted} uses smart contracts on the Ethereum blockchain either as a choreography monitor, where the smart contract monitors execution status and validity of workflow messages against a process model, or as an active mediator, where the smart contract''drives'' the process by sending and receiving messages according to a process model. BPMN models are translated into smart contracts. Local Ethereum nodes monitor the blockchain for relevant messages from the smart contract and create messages for the smart contract. Transaction cost and latency are recognized as important considerations in the evaluation of the approach. A comparison between the public Ethereum blockchain and the Amazon Simple Workflow Service cloud-based environment shows blockchain-based costs to be two orders of magnitude higher than a traditional infrastructure \cite{rimba2017comparing}. Hence, optimizing the space and computational requirements for smart contracts is important \cite{garcia2017optimized}. BPMN models are first translated to Petri Nets, for which minimizing algorithms are known. The minimized Petri nets are then compiled into smart contracts, achieving up to 25\% reduction in transaction cost \cite{weber2016using,weber2016untrusted}, while also significantly improving the throughput time. Building on lessons learned from \cite{weber2016using,weber2016untrusted}, Caterpillar is an open-source blockchain-based business process management system \cite{lopez2017caterpillar}. Developed in Node.js it uses standard Ethereum tools, like the Solidity compiler solc and the Ethereum client geth, to provide a distributed execution environment for BPMN-based process models. Lorikeet is a similar system \cite{DBLP:journals/insk/CiccioCDGLLMPTW19}, also based on BPMN models that are translated to smart contracts for the Ethereum chain. Also working with Ethereum and Solidity, \cite{STURM201919} present a system that focusing on resource management in addition to control flow considerations and extends the smart contracts to manage a variety of resource allocation patterns.

The replicated nature of blockchains means that information is available to all participants. One approach to address this privacy issue in the context of workflow management is the use of access control lists and their enforcement in smart contracts \cite{DBLP:conf/ithings/PourheidariRD18}.

\subsection{BFT Ordering in Blockchains}

After examining different blockchain consensus mechanisms in terms of termination time and fault tolerance, BFT-based consensus is recommended for business process executions \cite{ViriyasitavatHoonsopon2018blockchain}.

Solving the ordering and consensus problems not with expensive proof-of-work approaches, but with efficient and provably correct and live algorithms, is an important motivator for many recent blockchain projects. The Hyperledger project of the Linux foundation is the umbrella for a number of BFT-based blockchain implementations of various stages or maturity. Hyperledger Burrow\footnote{\url{https://www.hyperledger.org/projects/hyperledger-burrow}} is a blockchain that can execute Ethereum virtual machine code but is based on the Tendermint\footnote{\url{https://tendermint.com/}} BFT-based consensus algorithm. Hyperledger Iroha\footnote{\url{https://www.hyperledger.org/projects/iroha}} is based on ''YAC'', a proprietary BFT-based consensus protocol, but does not provide smart contracts. Hyperledger Indy\footnote{\url{https://www.hyperledger.org/projects/hyperledger-indy}} is a blockchain implementation for decentralized identity management, based on redundant byzantine fault tolerance (RBFT) \cite{DBLP:conf/icdcs/AublinMQ13}. Hyperledger Fabric\footnote{\url{https://www.hyperledger.org/projects/fabric}} is a generic blockchain implementation that provides smart contracts, called ''chaincode'', which can be written in Go or Node.js. Early implementations used the BFT-SMRT ordering protocol \cite{DBLP:conf/dsn/SousaBV18}, while recent versions have moved to the simpler, crash-fault tolerant (CFT) RAFT algorithm \cite{DBLP:conf/usenix/OngaroO14}.

\section{Blockchains}
\label{sec:blockchains}

A blockchain records transactions in contiguous blocks. A transaction can be any kind of content. Information integrity is maintained by applying a hash function to the content of each block, which also contains the hash of the previous block in the chain. Hence, altering a block requires changing all following blocks. In a typical blockchain, nodes are connected using a peer-to-peer network topology. New transactions may originate on any peer and must be recorded in new blocks. Blocks are generally distributed to each peer for independent validation and replicated storage. The key challenge is to achieve a consensus on the validity and order of transactions and blocks, despite peers that are characterized by ''byzantine faults'': they may not respond correctly, may respond unpredictably, or may become altogether unresponsive.

\subsection{Public and Permissioned Blockchains}

Blockchains may be either public or permissioned (''consortium''). Public blockchains typically have no access control or identity management. Hence, no node can be assumed to be trustworthy. In contrast, a permissioned blockchain has access controls, node operators are generally known and invited to participate, and (some) node operators may be implicitly trusted. The distinction between public and permissioned is not binary, but a continuum \cite{ViriyasitavatHoonsopon2018blockchain}.

Public chains are typically created to serve a large number of anonymous participants. Their advantages include anonymity, universal access, and generally a high trustworthiness as a large number of nodes provide independent transaction validation. On the other hand, public chains require incentives for validation, often in the form of a cryptocurrency, which increases transaction costs. Public chains also provide little flexibility to adapt to special use cases.

In contrast, permissioned chains are typically created for a specific use case with a small number of known institutional participants. Advantages of permissioned chains include low transaction costs, high flexibility to adapt to special use cases, identifiability of transaction originators, and access controls. Disadvantages may include relatively lower trustworthiness due to the smaller number of validating nodes.

Workflow management is typically the domain of a small number of institutional collaborators, rather than a large number of anonymous participants. As such, it is a good fit with permissioned blockchains.

While the blockchain technology used for public blockchains may also be used for permissioned blockchains, the different characteristics of the latter may permit or favour the use of technology options that would not be suitable for public blockchains, such as communication intensive BFT-based systems.



\subsection{Smart Contracts versus Application Code}

Smart contracts allow code execution as part of transactions on the blockchain. Advantages include code integrity, as code is part of the blockchain, and a tight integration of application logic with transaction validation. Disadvantages may be limitations of the smart contract language instruction set and the need to re-develop existing application logic.

In contrast, implementing application logic off-chain means that existing applications do not need to be ported, and developers have access to familiar programming languages, code libraries and development tools. On the other hand, transaction validation must call back to the application logic.

Smart contracts ensure that all nodes provide the same validation results, whereas performing validation in off-chain logic places the onus on the developers to ensure identical results for all nodes. On the other hand, it allows developers to develop against a behavioural specification without specifying the exact algorithms or implementation to be used. For the WfMS case in this article, that means that transparency is lost about the specific details of the workflow implementation, but what is gained is that different workflow systems can interoperate as long as all obey the same workflow semantics. 

Smart contracts have great potential in the context of workflow management, as witnessed by the the Caterpillar and Lorikeet approaches \cite{DBLP:journals/insk/CiccioCDGLLMPTW19}. However, neither Caterpillar nor Lorikeet provide a BPMN based generic workflow engine as a smart contract. Instead, both systems compile individual BPMN to specific smart contracts. \emph{Given the extensive investment in WfMS by researchers and practitioners, we believe that investigating how standard WfMS can be implemented on blockchain infrastructure \emph{without re-implementation} in smart contract languages is worthwhile.}

\subsection{Proof-of-Work Consensus}

Bitcoin popularized the proof-of-work mechanism for consensus finding and securing the blockchain. New transactions are distributed to all peers, validated and added to a transaction pool. Validation is based on transactions that exist in the chain as well as others already in the transaction pool. Each peer can independently propose new blocks based on its latest block and distribute these to other peers. Depending on network connectivity, speeds, and topology, each peer may have a different set of blocks and transactions, and hence may propose different blocks, leading to \emph{side branches}. Each peer considers the longest branch as the current main branch and proposes new blocks based on this. Transactions in side branches are not considered valid and are not considered when validating new transitions or blocks. When a side branch becomes longer than the current main branch, the chain undergoes a \emph{reorganization}. What was the side branch is validated and becomes the main branch. What was the main branch is considered invalid and becomes a side branch. Transactions no longer in the main branch are added back to the transaction pool to be included in other blocks. As a consequence, different peers can at times consider different blocks and transactions as valid. As proposed blocks are distributed across the network, peers will eventually converge on a consensus regarding the valid blocks and transactions and their order in the main branch of the chain.

To limit the rate of new block proposals and to secure the blockchain against atttacks, proof-of-work consensus requires block proposers to solve a hard problem (''proof-of-work'', ''mining''). Typically, this is to require the block hash to be less than a certain value. A limited block rate allows nodes to achieve eventual consensus, and a hard problem prevents attackers from ''overtaking'' the creation of legitimate blocks with fraudulent one. Assuming equal processing power for each node, the network needs $2f+1$ total nodes to tolerate $f$ faulty or malicious nodes.

The probability that a transaction in the main branch of the blockchain becomes invalid decreases with each block that is ''mined'' on top of it, although in principle it is always possible that a block becomes invalidated. Blockchain communities use rules of thumb for the number of additional blocks that is considered to make a transaction ''safe'' enough to act on. In addition to the lack of finality of consensus, this approach induces significant latency as applications must wait not only for one block but many to be created. Furthermore, applications must actively monitor the status of all transactions of interest, must react to chain reorganizations, and communicate these aspects to the user.

\subsection{BFT-Based Consensus and State Machine Replication}
\label{sec:bft}

In response to the drawbacks of the proof-of-work consensus, i.e. latencies, no finality of consensus, and required processing power, provably correct ordering algorithms, based on distributed systems research, have seen a resurgence in interest. Most of the ongoing research can be traced back to a practical method for achieving byzantine fault tolerance (PBFT) \cite{DBLP:journals/tocs/CastroL02}. PBFT orders client requests using a set of nodes that are fully connected by reliable messaging. Every ordering consensus is established by a specific set of nodes (''view''), with a leader or primary node. Tolerating up to $f$ faulty nodes requires $3f+1$ total nodes.

\paragraph*{Protocol}

PBFT is a three-stage protocol. A client sends a request to all nodes. The leader proposes a sequence number for the request and broadcasts a \emph{pre-prepare} message. Upon receipt of a \emph{pre-prepare} message, a node broadcasts a corresponding \emph{prepare} messge if it has itself received the request, has not already received another pre-prepare message for the same sequence number, and is in the current view. This indicates the node is prepared to accept the proposed sequence number. Nodes then wait to receive $2f$ matching \emph{prepare} messages, indicating that $2f+1$ nodes are prepared to accept the proposed sequence number for the request. When a node has received $2f$ identical \emph{prepare} messages, it broadcasts a \emph{commit} message to all nodes. Each node then waits to reeive $2f$ identical \emph{commit} messages, indicating that $2f+1$ nodes have accepted the proposed sequence number for the request. Upon committing, the node executes the request and sends a \emph{reply} message to the client. The client in turn waits for $2f+1$ identical replies, which indicates that a consensus has been reached on the sequence number of the request. 

In case the leader fails to propose a sequence number, nodes first forward requests to the leader. When the leader continues failing to act on requests or proposes sequence numbers too high or too low, nodes trigger a view change. The view change uses a three-stage protocol similar to the normal operation one to determine a new leader.

Consensus about request sequencing is closely related to state machine replication (SMR). Each node maintains a state that can be changed by client requests. When every node begins with the same state and executes requests in the same order, the state machine is replicated. 

\paragraph*{BFT SMART}

BFT-SMART \cite{DBLP:conf/dsn/BessaniSA14} is a software library built around the PBFT ordering protocol and adds dynamic view reconfiguration allowing nodes to join and leave views, and the MOD-SMART \cite{DBLP:conf/edcc/SousaB12} state transfer system.

Collaborative state transfer is useful when nodes create state checkpoints at different times (''sequential checkpointing''). Due to the lack of multiple identical checkpoints, a simple quorum protocol cannot be used. Instead, ''collaborative state transfer'' \cite{DBLP:conf/usenix/BessaniSFNC13} provides checkpoint and log information from multiple nodes in a way that allows a new node to verify its correctness.

BFT-SMART provides a simple programming interface. The client-side interface exposes the ability to submit requests for ordered or unordered operations. State-changing operations should be ordered, while read-only operations may be unordered. Applications implement a server-side interface, encapsulating the state machine, that receives ordered and unordered operation requests in consensus sequence from the BFT-SMART library for execution. Any replies are sent back to the requesting client. Operation requests are opaque to the library and are simple byte arrays. It is the client- and server-side application's responsibility to serialize and deserialize these in a meaningul way. View reconfigurations (adding or removing a node, or changing the level of byzantine fault tolerance) are special types of ordered requests but are treated as any other ordered request for ordering and consensus purposes.

For state management, the server-side application implements methods to fetch and set a state snapshot or checkpoint, also serialized as a byte array. State changes (ordered operations) are logged and the state is periodically checkpointed (sequential checkpointing). When a node joins a view, it is sent the latest checkpointed state (collaborative state transfer), which it sets for the server-side application, and any ordered operations after that checkpoint are then replayed, allowing the server state to catch up to the consensus state.

BFT-SMART has been proven to be correct and live, i.e. it will provide the same sequence of operations to all nodes and will not deadlock \cite{DBLP:conf/dsn/BessaniSA14}. In terms of throughput, a BFT-SMART system with four nodes ($f=1$) supports more than 15,000 operation requests (1kB size) per second with latencies around 10 milliseconds on a local network. BFT-SMART's performance decreases linearly as fault tolerance (and hence the number of nodes) increases: A system with 10 nodes ($f=3$) still supports more than 10,000 operations per second \cite{DBLP:conf/dsn/BessaniSA14}. 

\paragraph*{Summary}

PBFT-based ordering, as implemented in BFT-SMART, avoids the latency, lack of finality and processing requirements of proof-of-work consensus. On the other hand, its three-stage protocol imposes significant communication overhead and requires fully-connected nodes. Fault tolerance in PBFT-derived methods increases linearly with the number of nodes, but performance tends to decrease due to additional communication. \cite{vukolic2015quest} presents a comparison of proof-of-work and BFT consensus, shown in Table~\ref{tab:powbftcomparison}. \emph{The different strengths and weaknesses of the two consensus mechanisms suggest that BFT-based ordering is a good fit with small, permissioned blockchains in the workflow management context.}

\begin{table*}
\begin{center}
\begin{tabular}{|l|c|c|} \hline
 & {\bf Proof-of-work} & {\bf BFT ordering} \\ \hline
Node identity & open, anonymous & permissioned, nodes know other nodes \\ \hline
Consensus finality & no & yes \\ \hline
Scalability (ordering nodes) & excellent & limited \\ \hline
Scalability (clients) & excellent & excellent \\ \hline
Throughput & limited & excellent \\ \hline
Latency & high & low \\ \hline
Correctness proof & no & yes \\ \hline
\end{tabular}
\caption{Comparison between proof-of-work and BFT-based blockchains, adapted from \cite{vukolic2015quest}}
\label{tab:powbftcomparison}
\end{center}
\end{table*}

Note that while throughput (transactions per seconds) is a key performance metric for many blockchains and consensus algorithms, it is not important in the workflow management context: Even the largest organizations are unlikely to have production workflow systems that need to sustain tens of thousands of workflow actions per second. 

\section{Architecture and Design Choices}
\label{sec:architecture}

The main component of a WfMS is the workflow engine, which interprets the workflow model and enables work items for manual execution or execution by external applications \cite{hollingsworth1995workflow}. The engine maintains workflow state information and case data. It may be supported by, or include, services for organizational data management and role resolution, worklist management, document storage, etc. Designing a WfMS architecture requires choosing where to locate and how to implement the workflow engine and other service. 

Existing work on blockchain-based workflow management (Sec.~\ref{sec:related}) has deployed the workflow engine on the blockchain itself. However, by compiling a workflow model to a smart contract, the contract forms a workflow engine for only that workflow model. Alternatively, blockchains can be treated as a trusted infrastructure layer for generic workflow engines, using the blockchain only for storing and sharing the state of work and achieving consensus on that state. To our knowledge, there has been no such implementation using PBFT-derived, or any other, ordering mechanisms.

\begin{figure*}
\includegraphics[width=\textwidth]{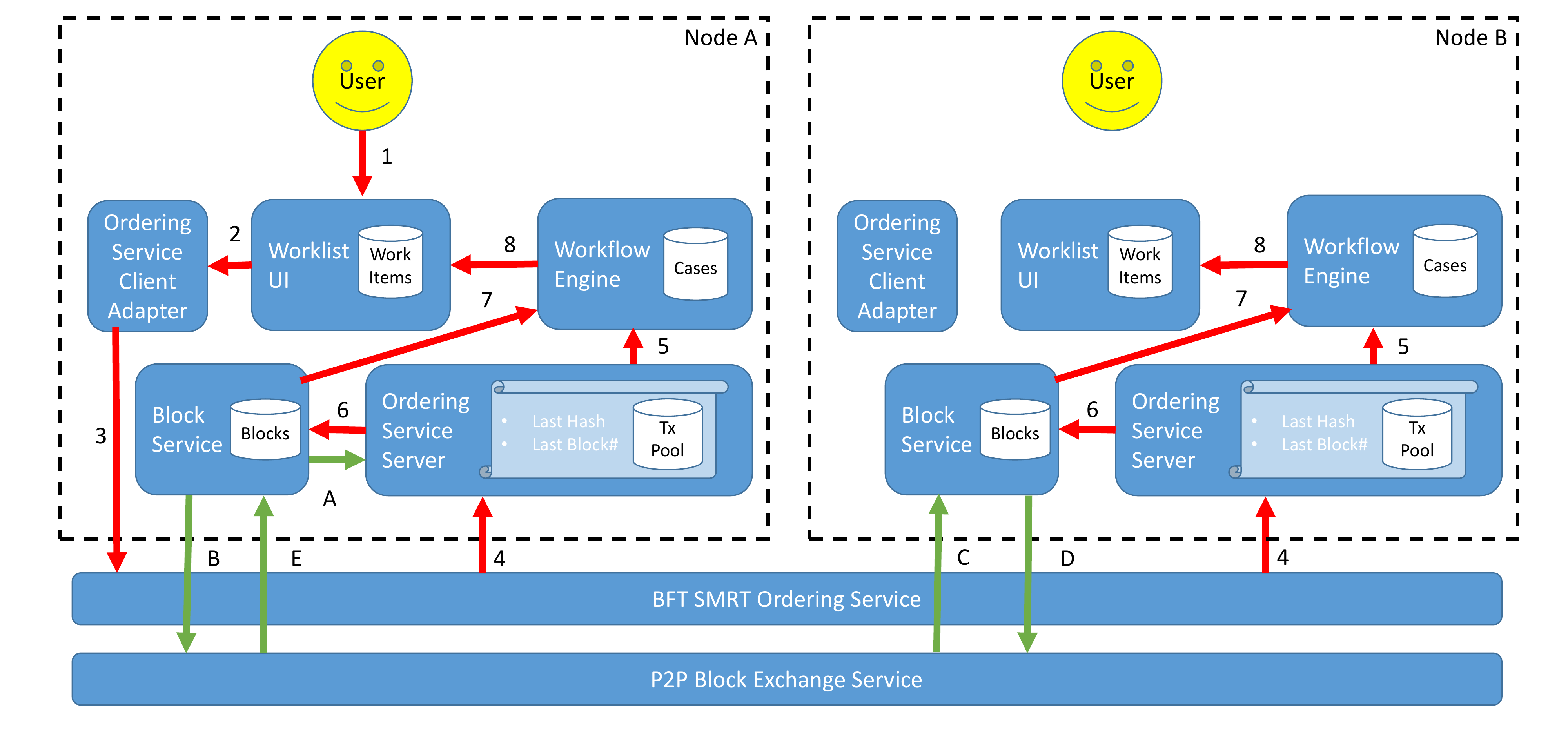}
\caption{Architecture overview, transaction flow, and block exchange}
\label{fig:architecture}
\end{figure*}

Ordering, block management, and the workflow engine are the three main services in our system architecture. Fig.~\ref{fig:architecture} shows the architecture of our system.

\paragraph*{Ordering Service} The ordering service in our prototype is implemented based on the BFT-SMART library \cite{DBLP:conf/dsn/BessaniSA14}. It can receive transactions to add to the blockchain, which is an ordered (state-changing) type of request it supports. The ordering service maintains a record of the latest block hash and block number, as well as a queue of transactions that have been added as its state. When a sufficient number of transactions has been collected, the ordering service creates a new block and clears the transaction queue. The ordering service returns the latest block hash and the hash of the set of queued transactions as a result to clients, allowing clients to detect absence of consensus. Clients can request the latest block hash.

\paragraph*{Block Service} The block service stores the blockchain, may exchange blocks with other nodes, and verifies the integrity of the blockchain. 

The block service uses a peer-to-peer network for block exchange with new and recovering nodes. This network is distinct from the network layer of BFT-SMART and is not fully connected. Block exchange is required only when a node begins operation and enters an ordering view. At that point, the ordering service state is first updated through the BFT-SMART state replication mechanisms. The block service then compares its latest block to the latest hash from the ordering service. The latter is assumed to be authoritative. Verification of the blockchain then proceeds backwards from the head of the chain, i.e. the block with the latest hash. Any missing blocks are requested from other peers and verified prior to adding them.

\paragraph*{Workflow Engine} The workflow engine maintains information about workflow instances (cases) and workflow model definitions. It receives workflow transactions of new blocks that are added to the chain, updating the state of each process instance and creating work items accordingly. Through the worklist, it manages user interactions with work items and execution of external functions by work items.

The red arrows labelled with numerals in Fig.~\ref{fig:architecture} indicate the steps of handling a workflow transaction in our system:

\begin{enumerate}
\item User completes work item in worklist
\item Transaction is created and passed to ordering service client for submission to ordering service
\item Transaction is submitted to ordering service
\item Transaction is passed in order to the ordering service server of all nodes
\item Ordering service servers validate ordered transaction with their workflow engine
\item When transaction pool contains a sufficient number of transactions, a new block is created and passed to block service
\item Block service notifies workflow engine of new block and transactions
\item Workflow engine updates state of running cases and creates new work items for local worklist 
\end{enumerate}

The green arrows labelled with letters in Fig.~\ref{fig:architecture} indicate the block exchange mechanism when a peer node is started. 

\begin{enumerate}
\renewcommand{\theenumi}{\Alph{enumi}}
\item Block service queries ordering service server for latest hash and transaction number
\item If block service determines it is missing blocks, it broadcasts a block request to all other nodes
\item Block services receive block requests
\item Block services assemble blocks into response message
\item Block service receives requested blocks and verifies block chain
\end{enumerate}

In step A, note that ordering service is started before block service and receives latest hash and transaction number through state exchange from other nodes. Furthermore, the block request contains the lower and upper block numbers required by the node. In step B, the block service begins by querying one random peer. When it receives no response, it queries an increasingly larger number of peers for blocks. In step D, other nodes only respond if they can satisfy at least the upper block number. In step E, if the block chain contains the most recent block but is missing individual earlier blocks, the block service will successively request these blocks from the peer it has most recently received blocks from. If this fails, it will again broadcast a query for a specific block. As fragments of the blockchain and individual blocks are added, the block services successively verifies the chain integrity beginning with the latest block and the last hash received from the ordering service.

Next, we discuss the architectural options that we considered when designing our prototype system. These affect performance, ease of implementation, and resilience. 

\subsection{BFT SMR State}
\label{sec:state}

Because BFT-SMART provides exchange of state information with new and recovering nodes, one architectural option is to employ this method also for the blocks of the blockchain. This means that the entire blockchain is part of the replicated state in BFT-SMART, effectively removing the need for a separate block service with its peer-to-peer network and block exchange protocol. While easy to implement by serializing the blockchain into the BFT-SMART state snapshot, this model becomes infeasible as the blockchain becomes too large to be rapidly exchanged with other nodes using the complex and communication-intensive collaborative state transfer mechanism in BFT-SMART. As an alternative, it is sufficient for the state to only contain the hash of the last block, the number of the last created block, and the queue of transactions waiting to be collected into new blocks.

\subsection{Block Creation}
\label{sec:blockcreation}

As noted above, blocks are created by the ordering service. One design option is to pass new blocks as replies from the ordering service operation back to the node that requested the add-transition operation that triggered the block creation. That node's block service is then responsible for exchanging the block with other nodes using the peer-to-peer network. This creates significant traffic on that network and may also lead to delays in new block distribution. 

A second design option, implemented in our system, is to have the ordering service server-side application that creates the new block pass the new block directly to the block service on its node. This tighter coupling between ordering service and block service reduces the communication overhead for the peer-to-peer network and latencies due to the block exchange. The peer-to-peer network is still required for block exchange with new or recovering nodes.

\subsection{Coupling of Block Service and Workflow Engine}
\label{sec:coupling1}

One option is for workflow engine and block service to always be present together on each node, as we have done for our system. Block service notifying the engine of new blocks, or the engine validating transactions for the ordering service can be done with local method calls.

While there is little to be gained by separating block service and workflow engine and running multiples of each, a second option is to operate only a single block service with multiple, distributed workflow engines. This eliminates the peer-to-peer network and block exchange communication. Blockchain integrity can still be verified from the latest hash of the ordering service nodes. However, this design eliminates the redundant storage that is an advantage of a replicated blockchain. On the other hand, redundancy can be achieved by a replicated storage layer within the block service, e.g. a distributed file system or database.

\subsection{Workflow State or Workflow Operations}
\label{sec:stateortransition}

A transaction may represent \emph{workflow operations} such as defining a new workflow model, launching a new case, executing an activity, aborting or cancelling a case or removing a workflow model. Activity execution information includes the activity name and case ID, as well input and output data values. Alternatively, a transaction can represent a \emph{workflow instance state}, i.e. data values and enabled activities, without capturing activity execution itself.

The first option requires the engine to maintain its own state of the workflow (i.e. information about workflow models, running instances, data values and enabled activities). Constructing this state means reading the blockchain \emph{forwards} from the genesis block and replaying all transactions. State updates are done by executing transactions in new blocks. While reducing the amount of information stored on the blockchain, as only changed information recorded, this option requires significant effort in managing the separate state and ensuring it is consistent with the blockchain record. In contrast, the second option makes the workflow state available by reading the blockchain \emph{backwards} from the head to identify the latest state for each process instance. State updates are done simply by copying workflow states from blockchain transactions as new blocks are presented. Not maintaining separate state signifanctly simplifies the workflow engine design but leads to more information being stored on the chain.

The first option provides activity information in each transaction. Hence, data constraints can be specified as post-execution constraints and checked when validating the transaction. The second option does not provide information about activity execution in a transaction. Hence, only global case data constraints can be specified and checked as part of transaction validation.

Finally, while transactions are waiting to be included in a block, users can be made aware of such pending transactions. For the first option, transactions are informative as they inform the user about pending workflow activities. In the second option, such transactions are less informative to the user, as they do not contain activity execution information.

\subsection{Block Size}
\label{sec:blocksize}

In proof-of-work blockchains, blocks contain multiple transactions. The block size is a trade-off among transaction arrival rate, available hashing power, desired block creation rate, available network bandwidth, and tolerance for latency. A transaction may be ''pending'' for a some time until it is included in a block and at a ''safe'' depth. In contrast, in BFT-based systems, there is no reason to prevent blocks from containing only one transaction, i.e. the blockchain becomes a chain of transactions. 

Moving to a chain of transaction has another advantage. Proof-of-work systems order transactions between different blocks, but the order of transactions within a block is not defined: Transactions may be included in the same block as long as they are not mutually contradictory. Block miners ultimately impose an order, but this order is arbitrary. This means that as pending transactions are collected, they must be validated against the \emph{entire} set of pending transactions to ensure they are not mutually conflicting. In a chain of transactions, a new transaction must be validated only against the \emph{immediately prior} one. 

\subsection{Coupling of Block Service and Ordering Service}
\label{sec:coupling2}

The ordering and block services (the latter always together with a workflow engine) can be coupled to varying degrees. At one extreme, block management is part of the ordering service, as discussed in Sec.~\ref{sec:state}.

In the less integrated architecture implemented in our system, every block service and workflow engine node is also an ordering node and vice versa, but block management is distinct from ordering and implements its own peer-to-peer network infrastructure. This allows each ordering node to quickly validate transactions using the local workflow engine. The drawback of this design is that the number of ordering nodes should be determined by the desired level of fault tolerance, whereas the number of workflow nodes should be determined based on the business process and/or application. An application requiring more ordering than workflow nodes is not a problem as the additional nodes are simply not assigned any workflow tasks. On the other hand, when an application requires more workflow nodes than ordering nodes, the excess ordering nodes decrease performance due to the communication overhead.

Both types of coupling have the problem that a faulty ordering service also compromises the block service and with that the workflow engine on that node. However, workflow engine and block service can detect their local node's faults when adding a transaction by comparing the consensus ordering service result to the result of the local ordering service result. If a detected fault is accidental, the node can be reset and synchronized with the consensus ordering view and receive valid blocks from peers. If the faulty behaviour is malicious, it is the intention of the process participant that controls the entire node, so the block service and workflow service are also compromised intentionally.

At the other extreme, in a very loosely coupled architecture, ordering nodes and block service / workflow nodes are separated. Newly created blocks are passed to the block server as replies from BFT operations and are communicated using the block service peer-to-peer network. However, because the ordering service validates transactions after ordering but before accepting them, each ordering node would require a reliable connection to at least one workflow engine. Managing these connections as workflow engines join and leave the network, and managing the additional communication, adds significant complexity and introduces additional latency in validating transactions. 

\section{Prototype Implementation}
\label{sec:prototype}

\begin{figure*}
\centering
\includegraphics[scale=0.5]{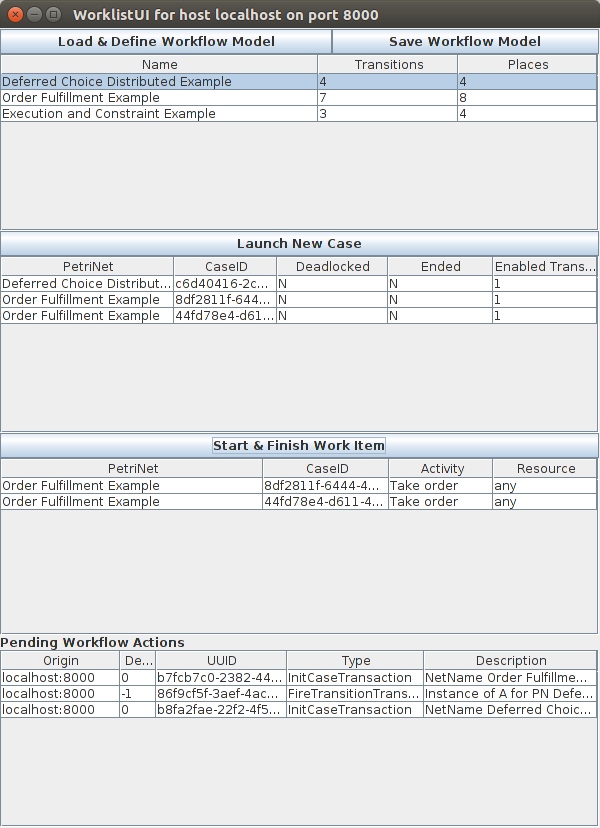}
\caption{Screenshot of prototype} 
\label{fig:screenshot}
\end{figure*}

Given the architectural design options discussed in the previous section and their advantages and disadvantages, we chose to implement our initial prototype by storing only the latest block hash, block number, and transaction pool as BFT-SMART state (Sec.~\ref{sec:state}). The workflow engine and block service are always present together at each node (Sec.~\ref{sec:coupling1}) and both are always co-located with an ordering service node (Sec.~\ref{sec:coupling2}). The ordering service passes new blocks directly to the local block service upon block creation, but a peer-to-peer network supports block exchange with new or recovering nodes (Sec.~\ref{sec:blockcreation}). We store workflow states on the blockchain, instead of workflow operations (Sec.~\ref{sec:stateortransition}) so as not having to maintain a separate workflow state in the engine. The block size is user configurable (Sec.~\ref{sec:blocksize}). We developed the prototype in Java. Source code is available\footnote{\url{https://joerg.evermann.ca/software.html}} as well as a video demonstration\footnote{\url{https://joerg.evermann.ca/BlockchainDemo.html}}. Fig.~\ref{fig:screenshot} shows a screenshot of our prototype.

We implemented a permissioned peer-to-peer infrastructure with a pre-defined list of participating actors. To keep our prototype simple, actors are identified by their internet address rather than their public keys, so that we can omit an address resolution layer. The P2P layer is implemented using Java sockets and serialization. Each P2P node has an outbound server that establishes connections to other peers, and an inbound server that accepts and verifies connection requests from peers. Each connection is served by a peer-connection thread, which in turn uses inbound and outbound queue handler threads to receive and send messages. Incoming messages are submitted to the inbound message handler which passes them to the appropriate service. Nodes can join and leave the peer-to-peer network at will. When a node joins, it tries to open connections to running peers. The first peer to be contacted will initiate a view change in the BFT-SMART odering service to include the new peer on that level as well. 

Upon starting of a node, the BFT-SMART layer will first update state information from other nodes in the view. Next, the block service will identify missing blocks and request them from peers. Once the blockchain is complete and verified, the workflow engine reads the blockchain to get the latest state for each workflow instance. Peer-to-peer messages are cryptographically signed and verified upon receipt. Table~\ref{tab:messages} lists the message types on our peer-to-peer network.

\begin{table*}
\begin{tabularx}{\linewidth}{|l|X|} \hline
BlockRequest & Requests a block with a specific hash from one or more peers \\ \hline
BlockSend & Sends a block to one or more peers \\ \hline
BlockChainRequest & Requests multiple blocks within a hash range from one or more peers \\ \hline
BlockChainSend & Sends multiple blocks to one or more peers \\ \hline
\end{tabularx}
\caption{Message types}
\label{tab:messages}
\end{table*}

Our blockchain has two transaction types. A \emph{ModelUpdate} transaction installs a new workflow model definition. An \emph{InstanceState} transaction contains a state of a workflow instance. It is submitted after a new case has been launched or an activity instance has been executed. Extensions to terminate cases and invalidate model definitions are readily possible.

To keep our prototype simple, our workflow models are based on plain Petri nets \cite{van1998application}. Each Petri net transition specifies a workflow activity. The workflow engine keeps track of the Petri net markings and case data, and can detect deadlocked and finished cases to remove them from the worklist. 

Each activity is associated with a single node. This partitioning of the process to different nodes does not form the resource perspective of the workflow but is used only to signal each node whether to act on a transaction. Each node can provide its own resource management by defining roles or other organizational concepts and performing further work item allocation within each node. Our models allow the process designer to specify this information. 

External method calls are specified as calls to static Java methods, and are performed synchronously by the workflow engine on work item enablement.

The data perspective is implemented as a key--value store. We currently admit only simple Java types as we implement a GUI for these; an extension to arbitrary types is readily possible. Each workflow instance has a set of data variables. When a transition is enabled, an activity instance (work item) is created for it and its input values are filled from the values of the workflow instance. The activity instance is then added to the local worklist or externally executed. After an activity instance is completed (manually or through execution of an external application), output values are written back to the workflow instance which is then submitted as an \emph{InstanceState} transaction to the ordering service.

We emphasize that our implementation is not meant to be a fully-featured WfMS. Instead, it serves only to illustrate generic WfMS functionality and its interplay with blockchain infrastructure components. The WfMS features themselves are not the focus of this research. 

The ordering service, workflow engine and the block service have a simple interface (Table~\ref{tab:interface1}). The ordering and block services can call on the workflow engine to validate transactions against the current workflow state, and optionally, against the most recent pending transaction. Validation checks that a transaction's instance marking is reachable from the marking of the current workflow instance state or that of the pending transaction. It also checks for data constraint violation. The block service receives new blocks from the ordering service and passes them to the workflow engine. In the other direction, the workflow engine can submit new transactions to the ordering service after a work item has been completed. Finally, the block service can request the latest hash from the ordering service on joining the network or recovering from a fault. 

\begin{table*}
\begin{tabularx}{\linewidth}{|l|l|X|} \hline
$\rightarrow$ & validateTransaction(tx[, pendingTx]) & Ordering service asks workflow engine to validate a transaction, given the current workflow state and optionally the most recent pending transaction (cf. arrow 5 in Fig.~\ref{fig:architecture}) \\ \hline
$\rightarrow$ & receiveBlock(block) & Block service receives a new block and passes relevant transactions to the workflow engine (cf. arrow 6 in Fig.~\ref{fig:architecture}) \\ \hline
$\leftarrow$ & addTransaction(tx) & Workflow engine submits a new transaction to the ordering service (cf. arrow 2 in Fig.~\ref{fig:architecture}) \\ \hline
$\leftarrow$ & getLatestHash() & Block service requests the latest hash from the ordering service (cf. arrow A in Fig.~\ref{fig:architecture}) \\ \hline
\end{tabularx}
\caption{Interfaces between ordering service, block service and workflow engine (directions from the perspective of the ordering service)}
\label{tab:interface1}
\end{table*}

\section{Discussion and Conclusions}
\label{sec:discussion}

Previous work on blockchain-based WfMS has focused on creating smart contracts to represent specific workflow models. In particular, the Ethereum proof-of-work-based blockchain is widely used. However, proof-of-work-based systems have significant drawbacks in terms of processing power requirements, latency, and the lack of final consensus. In this work, we have shown that a PBFT-derived ordering and consensus method is a suitable WfMS infrastructure. While we do not use smart contracts of modern blockchains, the use of a blockchain remains essential, as it provides independent validation of workflow actions, distribution, replication, and tamper-proofing to workflow management systems.

Through the development of our prototype, we have identified architectural design options with their advantages and disadvantages. Our chosen design, in which we integrate ordering service, block service, and workflow engine on every node, strikes a balance between architectural and implementation simplicity on the one hand, and performance and scalability on the other.

A limitation in our chosen model is that the number of nodes must strike a balance between the requirements of the workflow (the number of actors involved), the desired level of fault tolerance, and the performance of the system. The major advantages are the low communication overhead on the P2P block exchange and the ability of local workflow engines to validate transactions quickly.

While our approach has lower resilience against faults and malicious attacks than proof-of-work chains, it also has lower latency and higher throughput. Unlike proof-of-work chains, the PBFT-based approach does not scale to a very large number of nodes. Given these characteristics, systems such as ours are suitable for permissioned blockchain applications. The low latency makes them suitable for fast-moving processes, where activities are of short duration or must follow each other quickly. Our system is cheaper to operate than public proof-of-work blockchains that incentivizes block mining through cryptocurrencies. While one can implement permissioned proof-of-work chains, these lose their resilience against attacks in small networks as it is easy for a single actor to acquire the majority of processing power in a single high-performance node. Attacking a PBFT-based system cannot be done by concentrating computational power but requires control of more than $1/3$ of all nodes, which is more difficult to achieve, especially in the absence of trust among actors. 

From the user's perspective, our system is little different from a traditional WfMS. Because transactions need not wait to be included in the blockchain or to be mined to a ''safe'' depth and consensus is final, latency is low, and the execution status of workflow activities cannot change and does not need to be monitored or reported to the user.

Our work on the prototype implementation has shown some avenues for future research.

\begin{itemize}
\item We currently assign single peer nodes statically to workflow activities. In the future, we will extend this to dynamic peer node assignment and to integrate this with the workflow's resource perspective.

\item Porting existing feature-complete workflow engines, such as the open-source YAWL \cite{ter2009modern} or Bonita\footnote{\url{https://www.bonitasoft.com}} systems, to blockchain infrastructure allows a richer workflow language and leverages existing implementations. 

\end{itemize}

To conclude, this paper has described a prototype implementation for an architecture that has not yet seen any attention in the blockchain-based workflow literature. We have implemented PBFT-based system as recommended by \cite{ViriyasitavatHoonsopon2018blockchain} and shown that this infrastructure is suitable for WfMS. We have shown how generic workflow engines can be readily adapted to fit onto a blockchain infrastructure without implementing these as smart contracts. The interfaces between components are quite simple. In contrast to \cite{mendling2018blockchains}, who suggest that blockchain-specific modelling languages need to be developed, our work shows that workflow engines do not need to be implemented using smart contracts, as done by \cite{weber2016using,weber2016untrusted}, but that traditional workflow engines be easily adapted to use blockchains as infrastructure for communication, persistence, replication, and trust building.

\bibliographystyle{splncs04}

\end{document}